\documentclass[12pt]{iopart}
\usepackage{graphicx}
\usepackage{multicol}
\usepackage{flafter}
\usepackage[left=1.7cm,right=1.6cm,top=2cm,bottom=2cm,a4paper]{geometry}
\usepackage{cuted}
\begin{document}
\twocolumn 
\begin{strip}
\begin{center}
{\Large \textbf{Impurity intrusion in radio-frequency micro-plasma jets operated in ambient air}}
\end{center}
\author{B Niermann, A Kanitz, M B\"oke and J Winter}

\address{$^1$Ruhr-Universit\"at Bochum, Institute for Experimental Physics II, Universit\"atsstra\ss e 150, 44780 Bochum, Germany}

\ead{benedikt.niermann@rub.de}

\begin{abstract}
Space and time resolved concentrations of helium He ($^3\mbox{S}_1$) metastable atoms in an atmospheric pressure radio-frequency micro-plasma jet were measured using tunable diode laser absorption spectroscopy. Spatial profiles as well as lifetime measurements show significant influences of air entering the discharge from the front nozzle and of impurities originating from the gas supply system. Quenching of metastables was used to deduce quantitative concentrations of intruding impurities. The impurity profile along the jet axis was determined from optical emission spectroscopy as well as their dependance on the feed gas flow through the jet.
\end{abstract}

\submitto{\JPD}

\maketitle

\end{strip}

\section{Introduction}

In recent years the research on atmospheric pressure micro-plasmas became a strong focus in plasma sciences, mostly due to their high potential for new plasma applications without the need for expensive vacuum equipment \cite{becker}. Among the large variety of micro-plasma sources, that make use of DC, pulsed DC and AC ranging from mains frequency to RF, atmospheric pressure plasma jets provide a simple design by featuring an $\alpha$-mode RF discharge between two bare metallic electrodes.

Jet discharges are particularly well suited for applications in ambient atmosphere, like biological, medical, or a variety of surface treatment applications. Operating the discharge in direct contact with air, however, makes the plasma vulnerable to intrusions of nitrogen and oxygen molecules from the surrounding environment. This may reduce the operational space, induce instabilities and create unwanted reactive species. To study these influences in detail, excited species in the discharge provide an indirect measure for the amount of impurities in the discharge, since they react very sensitively to collisions with these species. Especially metastable species carry large amounts of potential energy and are a source of ionization. In this context helium metastable species pose an appropriate indicator, since helium is a widely used carrier gas in atmospheric pressure micro-discharges. Compared to molecular gases, the excitation threshold of helium metastable atoms is high, and can be exceeded just by high energetic electrons in the tail of the EEDF. Thereby the metastable excitation process is highly sensitive to impurities in the discharge. Due to their long lifetime they collide more frequently with other particles (about 15000 times per $\mu$s at atmospheric pressure), which makes them important for plasma chemistry processes and very sensitive to even tiny amount of molecular gases in the discharge. Especially nitrogen and oxygen molecules have some of the largest quenching cross sections for metastable helium atoms \cite{pouvesle}. The metastable density in micro-discharges is several orders of magnitude lower than the density of the ground-state atoms. However, compared to most other species the density is significant and the electron collision excitation cross sections of some helium levels out of the metastable states exhibit values which are several orders of magnitude larger and have much lower thresholds than those for the ground state \cite{katsch,flohr}. Thus, metastables strongly contribute to the ionization.

A reliable technique for the systematic investigation of metastable species is tunable diode laser absorption spectroscopy (TDLAS), since it is not invasive and provides absolute population densities. We applied TDLAS to record the spectral profiles and the lifetime of the lowest helium metastable state, deducing absolute densities with high sensitivity and spatial resolution, revealing detailed information about the distribution of impurities in the discharge. In addition to TDLAS we applied optical emission spectroscopy (OES) to observe nitrogen emissions lines that provide direct information about the nitrogen content in the discharge.

\section{The atmospheric pressure micro-plasma jet}

\begin{figure}
    \centering
    \includegraphics[width=0.5\textwidth]{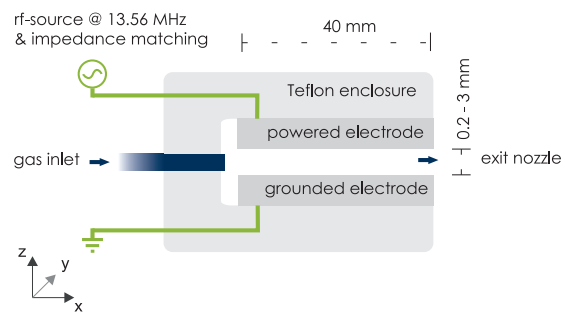}
        \caption{Sketch of the micro-plasma jet discharge. Shown is a 2-dimensional cross-section through plane that is spanned by the two electrodes.}
        \label{figure1}
\end{figure}

The atmospheric pressure micro-plasma jet is a capacitively coupled, non-\-thermal glow-discharge at high pressures. The design concept of this discharge is based on the plasma jet introduced by Selwyn et al. in 1998 \cite{selwyn} and advanced by Schulz-von der Gathen et al. \cite{schulz}. Feed gas flows between two closely spaced stainless steel electrodes driven at 13.56 MHz radio-frequency in a parallel plate configuration (Fig. \ref{figure1}). Electrodes and plasma volume are enclosed by quartz windows, giving direct optical access to the discharge. The jet is operated in helium at various flow rates from 200\,sccm to 5\,slm.  The electric field between the electrodes causes a breakdown in the gas and produces a plasma with electron temperatures and densities of about 1 to 2\,eV and $10^{10 }\mbox{cm}^{-3}$, respectively \cite{schaper,waskoenig}. Atoms and molecules in the feed gas become excited, dissociated or ionized by electron impacts. Since the electrons are not in thermal equilibrium with the ions and neutrals, the gas temperature remains a few tens of K above room temperature \cite{knake}. The distance between the two windows is fixed to 1\,mm while the electrode gap size is variable between 0.2 and 3\,mm. In all here presented configurations the discharge operates as a typical $\alpha$-mode rf glow discharge.

The whole jet setup is located in an airtight stainless steel vessel, whose atmosphere can be defined accurately. Thus, the jet can either be operated in ambient air, or in a pure helium environment, and a precise retracement of leak channels is possible.

\section{Spectroscopic setup}

For the absorption spectroscopic measurements a standard TDLAS setup was used. Details about the setup and procedure for calculating absolute metastable densities are published in \cite{niermann}. For the measurement of the absorption signal across the jet axes, the discharge casing was mounted on a small movable stage featuring three electronically controlled stepping motors to adjust, with high precision, the positions of the discharge in all spatial dimensions. This setup allows the positioning of the jet with an accuracy of about 5 $\mu$m and automated x-z-mapping of the complete plasma volume. In y-axis all measurements presented in this paper are line averaged.
OES measurements were carried out using a lens system to image the plasma emissions into a fiber, guiding the signal to a spectrometer with a spectral resolution of 0.75\,nm. Emission profiles are averaged in y and z dimensions. The spatial resolution in x is 1\,mm. The jet was operated in pulsed mode with a frequency of 4\,kHz and a duty cycle of 50\,\%, resulting in power on and off times of 125\,$\mu$s, 20 times longer than the maximum lifetime of the species.

\section{Results and discussion}

\subsection{Metastable quenching by air}

\begin{figure}
    \centering
    \includegraphics[width=0.5\textwidth]{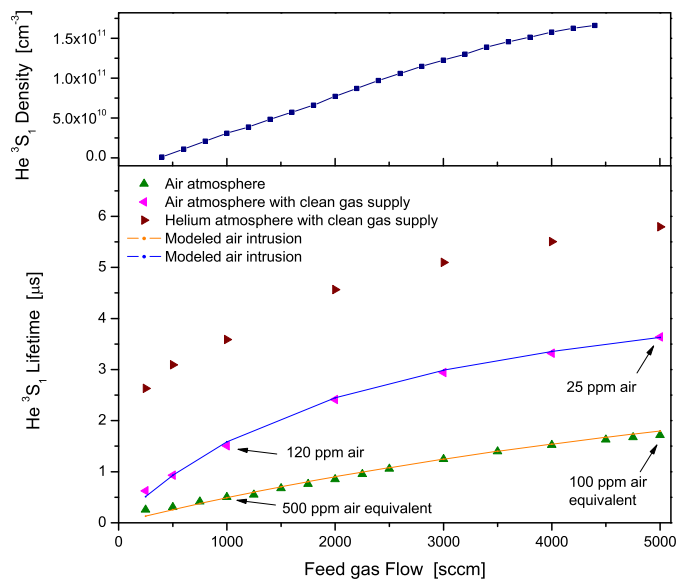}
        \caption{He ($^3\mbox{S}_1$) metastable density (top) and lifetime (bottom) for different setups in dependence on the feed gas flow. Also shown is the modeled lifetime taking into account significant contribution of nitrogen and oxygen quenching.}
        \label{figure2}
\end{figure}

The data shown in figure \ref{figure2} (upper graph) represent the metastable density in dependence on the feed gas flow. They have been taken in the center of the discharge (2\,cm from the exit nozzle) while the jet was running surrounded by ambient air. The density increases with the gas flow rate and shows a weak tendency to saturate at higher flows. This strong dependence can not be explained by a fundamental change in the excitation mechanism, since neither the pressure nor the outer power coupling to the discharge is changing. To determine the origin of this mechanism in detail the decay rate of the density in the afterglow was measured under different conditions.

Figure \ref{figure2} (bottom graph) indicates that the lifetime of the species increases also with the gas flow and, like the density, shows a tendency to saturate for higher flows. Assuming infinite purity of the helium gas and proposing that the metastable lifetime is mainly determined by the three-body collision process:
\begin{eqnarray} He^* + 2He \rightarrow He^*_2 + He,\end{eqnarray}
the theoretical lifetime of He* ($^3\mbox{S}_1$) in pure helium is expected to be
\begin{eqnarray}
\tau=(K_3\cdot N_{He}^2)^{-1}=5.8\,\mu \mbox{s},
\end{eqnarray}
with $K_3=2.5\cdot10^{-34}\,\mbox{cm}^6\mbox{s}^{-1}$ the rate coefficient for three-body collisions and $N_{He}=2.68\cdot10^{19}\mbox{cm}^{-3}$ the helium ground state density \cite{tachibana,ueno}.  The loss of metastable atoms by two-body collisions as well as by diffusion can be omitted since the diffusion loss frequency is 2 orders of magnitude lower then the loss frequency by three-body collisions. Three situations have to be distinguished in this context. In the first setup (red triangles in figure \ref{figure2}) the airtight vessel around the jet was evacuated and filled with pure helium. Furthermore the gas supply system, that is the stainless steel tube connection from the gas bottle to the micro-plasma jet, was kept under a vacuum of 1\,Pa for some time, to clean it from traces of water and air. As a result for high flows of 5\,slm the theoretical lifetime for a clean helium feed gas can be reached, but the lifetime is dropping significantly for lower flows. Taking into account that the leakage rate of the gas supply system can be neglected, the metastable quenching can be attributed to tiny amounts of water and air molecules desorbing from the surface and contaminating the feed gas in the order of a 2 to 40\,ppm. This fraction of contamination decreases linearly with the feed gas flow.

In a second setup (pink triangles in figure \ref{figure2}) the airtight vessel was filled with ambient air. The measured lifetimes in this case are significantly below the theoretical value for a clean feed gas. Again the discrepancy can be attributed to the loss of metastable atoms by molecular excitation processes. Due to the low ionization thresholds these excitation processes are mostly Penning ionizations, in this case, with air introduced through the front nozzle:
\begin{eqnarray} He^* + N_2 \rightarrow He + N_2^+ + e,\end{eqnarray}
\begin{eqnarray} He^* + O_2 \rightarrow He + O_2^+ + e.\end{eqnarray}
Assuming that the dominant impurity contribution is due to residual $N_2$ and $O_2$ molecules, the impurity level can be estimated by a simple model. Taking into account quenching collisions with nitrogen and oxygen molecules the lifetime is limited to
\begin{eqnarray}
\fl\tau=&[K_3\cdot N_{He}^2+K_{N_2}(N_{N_2,\mbox{Bottle}}+ N_{N_2}(\Gamma))\\
\fl&+K_{O_2}(N_{O_2,\mbox{Bottle}}+ N_{O_2}(\Gamma))]^{-1},
\end{eqnarray}
with $K_{N_2}=7\cdot 10^{-11}\mbox{cm}^{3}\mbox{s}^{-1}$ and $K_{O_2}=2.5\cdot 10^{-10}\mbox{cm}^{3}\mbox{s}^{-1}$ the rate constants for Penning ionization processes with molecular nitrogen and oxygen and $N_{N_2}(\Gamma)$ as well as $N_{O_2}(\Gamma)$ the flow dependant nitrogen and oxygen densities \cite{ueno,cardoso}. $N_{N_2,\mbox{Bottle}}$ and $N_{O_2,\mbox{Bottle}}$ are the base contaminations of the feed gas, given by the purity of the helium used in the experiment. This base contamination sums up to about 2\,ppm. Since we are assuming the intrusion of air into the system, the ratio between $N_2$ and $O_2$ is fixed to 3.7. Furthermore we presume the intrusion air to vary anti-proportionally with the flow rate. The result of the model is shown in figure \ref{figure2} (blue line). It agrees well with the measured values. We determine the level of air introduced through the nozzle to be in the order of a few dozen ppm (120\,ppm for 1\,slm flow, 25\,ppm for 5\,slm flow).

In a third setup (green triangles in figure \ref{figure2}) the micro-plasma jet as well as the gas supply system was exposed to ambient air. Repeating the flow dependant lifetime measurements the values show again a significant decrease. Nevertheless, the amount of air introduced into the system through the front nozzle must be the same like in the previous case. Additional metastable quenching can only be caused by traces of air and water desorbing from the surfaces of the gas supply system as well as from the metallic electrodes. This assumption is supported by mass spectrometry and OES measurements that show significant amount of water ions as well as nitrogen and oxygen emissions in the plasma and the effluent region. Matching these measurements with the previously described model (orange line) the observed quenching corresponds to a few hundred ppm of air molecules in the feed gas stream (500\,ppm for 1\,slm flow, 100\,ppm for 5\,slm flow).

\subsection{Longitudinal impurity profile}

\begin{figure}
    \centering
    \includegraphics[width=0.5\textwidth]{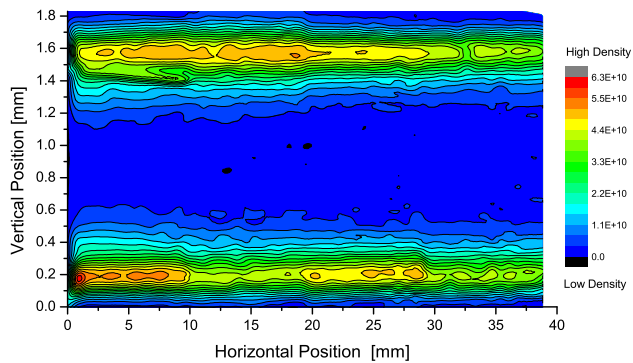}
        \caption{2-dimensional map of the measured metastable density in the discharge volume for the He $2^3\mbox{S}_1\rightarrow\,2^3\mbox{P}^0_{1,2}$ transition. Powered electrode at the top, grounded electrode at the bottom. Electrode gap size was 1.8\,mm. Densities are given in cm$^{-3}$. Measurements were taken in under ambient atmosphere.}
        \label{figure3}
\end{figure}

Figure \ref{figure3} shows a 2D-map of the He ($^3\mbox{S}_1$) metastable density in the discharge volume. Both, horizontal and vertical axis show the exact area between the electrodes. The map covers 2.000 reading points (40 vertical x 50 horizontal) of the absorption signal in the plasma volume. A variety of effects determine the metastable distribution in the vertical and in the longitudinal axis. While the vertical profile reflects the RF-sheaths structure and is determined by the electron density and temperature distribution, the longitudinal profile is governed mainly by environmental factors \cite{niermann}.

\begin{figure}
    \centering
    \includegraphics[width=0.5\textwidth]{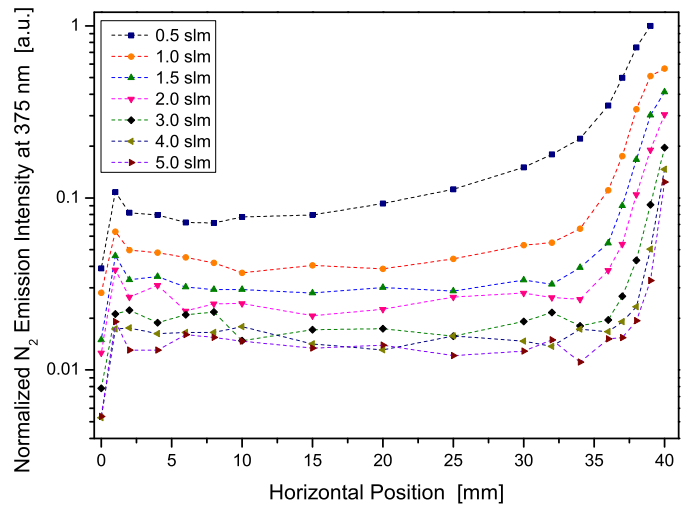}
        \caption{Longitudinal emission profile of the second positive system of molecular nitrogen at 357\,nm. Measurements were taken under ambient atmosphere.}
        \label{figure4}
\end{figure}

As discussed before the impurities entering the jet through the front nozzle are quenching the metastable atoms. Since the jet is running permanently in contact with the ambient air the intrusion of nitrogen and oxygen into the plasma channel is significantly high, and decreases the metastable density especially in the first millimeters from the nozzle. This observation is supported by OES measurements showing the emission of various nitrogen lines. Figure \ref{figure4} presents the longitudinal profile of the molecular nitrogen emissions at 357\,nm, representing the second positive system. The measurements reveal strong nitrogen emissions in the first 10\,mm from the front nozzle. The intrusion depth increases with decreasing gas velocities, ranging from about 10 to 100\,ms$^{-1}$. Especially for small gas flows a back diffusion of nitrogen can be observed over the whole length of the jet. The back diffusion of air into the system is promoted by the geometry of the gas channel, whose shape is rectangular. Especially in the corners of the channel the gas flow is likely not to be laminar and turbulences may occur.

\section{Summary}

It was shown that in atmospheric pressure micro-plasma jets operated in contact with ambient air the back diffusion of nitrogen and oxygen into the gas channel is a significant effect with strong impact on the energy transfer processes of the discharge. Furthermore the inner surfaces of the gas supply system and of the jet discharge itself pose a source of impurities contaminating the the feed gas. These influences and the likely consequences on the plasma chemistry have to be taken into account for any application that envisage the use of such discharges in ambient atmosphere.

\ack
This project is supported by DFG (German Research Foundation) within the framework of the Research Group FOR1123 and the Research Department 'Plasmas with Complex Interactions' at Ruhr-University Bochum.

\newpage
\clearpage


\end{document}